\documentclass[conference]{IEEEtran}
\IEEEoverridecommandlockouts
\usepackage{cite}
\usepackage{booktabs}
\usepackage{amsmath,amssymb,amsfonts}
\usepackage{algorithmic}
\usepackage{graphicx}
\usepackage{tcolorbox}

\usepackage{textcomp}
\usepackage{xcolor}
\usepackage[colorlinks=true,linkcolor=blue,citecolor=blue,urlcolor=blue]{hyperref}
\def\BibTeX{{\rm B\kern-.05em{\sc i\kern-.025em b}\kern-.08em
    T\kern-.1667em\lower.7ex\hbox{E}\kern-.125emX}}
\begin{document}

\title{TourSynbio-Search: A Large Language Model Driven Agent Framework for Unified Search Method for Protein Engineering
}

\author{
\IEEEauthorblockN{
Yungeng Liu\textsuperscript{1,2}, 
Zan Chen\textsuperscript{1},
Yu Guang Wang\textsuperscript{1,3},
Yiqing Shen\textsuperscript{1,4,*}
}
\IEEEauthorblockA{
\textsuperscript{1}\textit{Toursun Synbio}, Shanghai, China\\
\textsuperscript{2}\textit{Department of Computer Science}, \textit{City University of Hong Kong}, Hong Kong, China\\
\textsuperscript{3}\textit{Institute of Natural Sciences}, \textit{Shanghai Jiao Tong University}, Shanghai, China\\
\textsuperscript{4}\textit{Department of Computer Science}, \textit{Johns Hopkins University}, Baltimore, MD, USA\\
{\textsuperscript{*}Corresponding authors.}\\
{\footnotesize yiqingshen1@gmail.com}
}}

\maketitle

\begin{abstract}
The exponential growth in protein-related databases and scientific literature, combined with increasing demands for efficient biological information retrieval, has created an urgent need for unified and accessible search methods in protein engineering research. 
We present TourSynbio-Search, a novel bioinformatics search agent framework powered by the TourSynbio-7B protein multimodal large language model (LLM), designed to address the growing challenges of information retrieval across rapidly expanding protein databases and corresponding online research literature. 
The agent's dual-module architecture consists of PaperSearch and ProteinSearch components, enabling comprehensive exploration of both scientific literature and protein data across multiple biological databases.
At its core, TourSynbio-Search employs an intelligent agent system that interprets natural language queries, optimizes search parameters, and executes search operations across major platforms including UniProt, PDB, ArXiv, and BioRxiv
The agent's ability to process intuitive natural language queries reduces technical barriers, allowing researchers to efficiently access and analyze complex biological data without requiring extensive bioinformatics expertise.
Through detailed case studies in literature retrieval and protein structure visualization, we demonstrate TourSynbio-Search's effectiveness in streamlining biological information retrieval and enhancing research productivity. 
This framework represents an advancement in bridging the accessibility gap between complex biological databases and researchers, potentially accelerating progress in protein engineering applications.
Our codes are available at \url{https://github.com/tsynbio/Toursynbio-Search}.
\end{abstract}


\section{Introduction}
The exponential growth of protein-related data and scientific literature has created unprecedented challenges for researchers in accessing and synthesizing relevant information for protein engineering applications \cite{stephens2015big}. 
While traditional database interfaces offer comprehensive coverage, they often require specialized query syntax and navigation across multiple platforms, creating barriers to efficient information retrieval \cite{zou2015biological}. 
Although recent advances in retrieval-augmented generation (RAG) have demonstrated promising results in general information retrieval tasks, these approaches face unique challenges in the biological domain \cite{lewis2020retrieval}. 
Specifically, current RAG-based frameworks struggle with two key limitations, namely maintaining accuracy amid rapidly updating protein databases and research literature, and providing precise responses to complex biological queries that require domain-specific understanding \cite{zhu2023large}. 
Furthermore, existing general-purpose search frameworks lack the specialized understanding required for accurate protein-specific queries and often fail to provide precise feedback for complex biological information requests \cite{jensen2006literature}. 
These challenges highlight the need for an integrated search solution that combines advanced natural language processing capabilities with specialized biological data.

To address these limitations, we present TourSynbio-Search, a novel search agent framework built upon the TourSynbio-7B protein multimodal large language model (LLM) model \cite{shen2024toursynbio}.
TourSynbio-7B processes protein sequences directly as natural language, eliminating the need for external encoders and achieving state-of-the-art performance on protein-related understanding tasks \cite{gpt4, shen2024fine}. 
At the core of TourSynbio-Search lies a three-layer agent architecture that orchestrates intelligent search operations.
The architecture consists of an LLM-powered agent match layer for query interpretation, a parameter refinement layer for search optimization, and an execution layer that coordinates data retrieval across multiple sources. 
Moreover, to provide comprehensive access to biological information, we developed a dual-module search framework that seamlessly integrates literature and protein data retrieval \cite{wooldridge1999intelligent, chang2024survey}. 
The PaperSearch module facilitates scientific literature retrieval from ArXiv and BioRxiv repositories, while the ProteinSearch module enables efficient access to protein data from PDB and UniProt databases, enhanced by integrated PyMOL visualization capabilities. 
This unified approach eliminates the need for researchers to navigate multiple platforms independently.

The major contributions are three-fold.
\begin{itemize}
    \item We propose a three-layer search agent, namely the TourSynbio-Search.
    It comprises an agent match layer employing LLM for query analysis, a parameter supplement and confirmation layer for search optimization, and an execution layer coordinating data retrieval operations \cite{wooldridge1999intelligent, chang2024survey}.

    \item We propose a dual-module search framework that unifies biological information access, where PaperSearch handles scientific literature retrieval through ArXiv\cite{de2016arxiv} and BioRxiv \cite{sever2019biorxiv}, while ProteinSearch manages protein data acquisition from PDB \cite{burley2017protein} and UniProt \cite{uniprot2015uniprot} with integrated visualization capabilities \cite{delano2002pymol}.

    \item We propose an adaptive query processing mechanism that transforms natural language inputs into structured database queries.
    It can automatically supplement and verify search parameters, and adjust responses based on result quality and user feedback.
\end{itemize}


\begin{figure*}[htbp!]
\centering\centerline{\includegraphics[trim={0.0cm 0.cm 0.05cm 0.05cm},clip,width=1.0\linewidth]{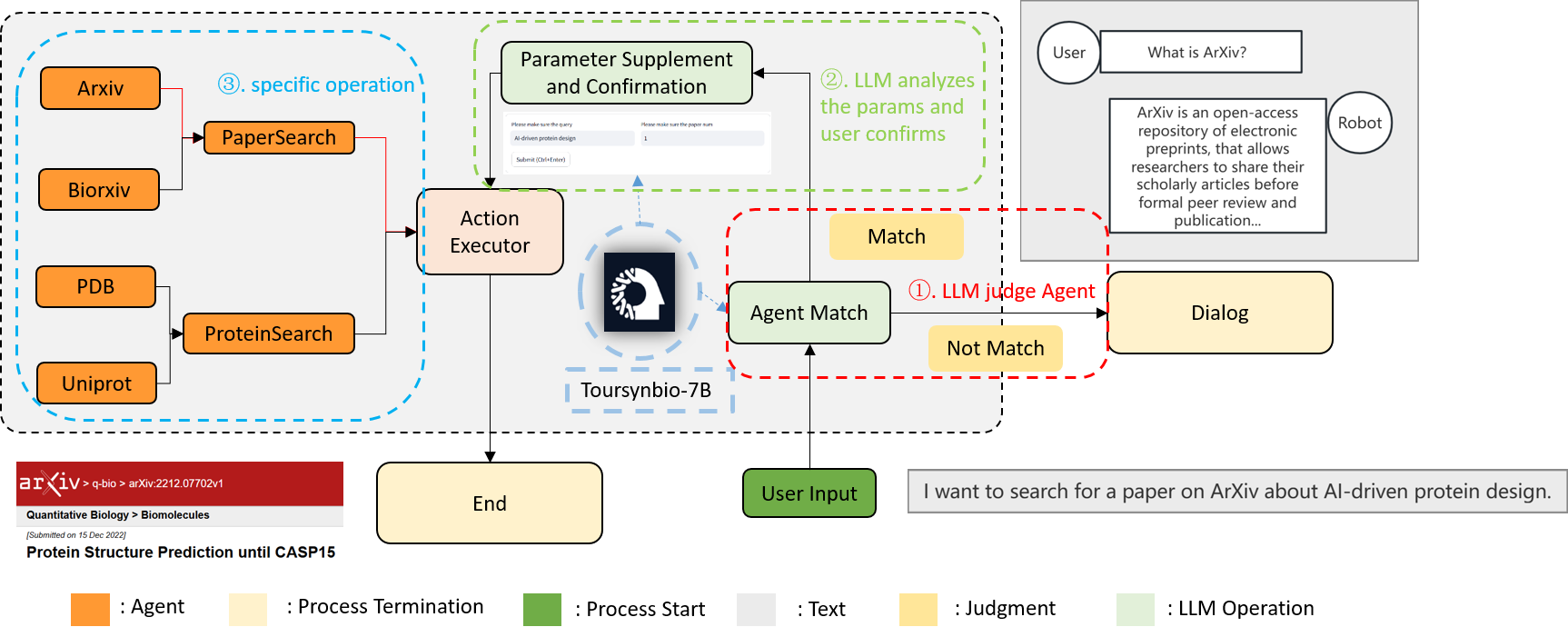}}
\caption{
Architectural overview of the TourSynbio-Search framework. 
It implements a three-layer agent architecture for information retrieval across biological databases. 
The agent match layer employs TourSynbio-7B to analyze user queries and determine appropriate agent routing. 
The parameter supplement and confirmation layer processes natural language inputs into structured database queries, allowing user verification of extracted parameters. 
The action executor layer coordinates data retrieval operations across multiple sources including arXiv and bioRxiv for scientific literature, and PDB and UniProt for protein data. 
If no agent match is found, the TourSynbio-Search defaults to direct dialogue mode with TourSynbio-7B. 
The framework integrates specialized search strategies for each data source while maintaining a unified user interface, as demonstrated by the example query shown for the LLM-driven protein design paper search.
}
\label{ex3}
\end{figure*}

\section{Methods}

\subsection{Toursynbio-7B}
At the core of our search framework lies Toursynbio-7B, a specialized protein-focused multimodal LLM built upon InternLM2-7B \cite{team2023internlm,shen2024toursynbio}. 
The model was fine-tuned on ProteinLMDataset, our curated collection encompassing 17.46 billion tokens of diverse protein-related content, including scientific literature, protein sequences, structural data, and functional annotations across multiple languages \cite{data}.
Toursynbio-7B's architecture enables the direct processing of protein sequences as a natural language without requiring external encoders or embeddings.
This unique capability facilitates the seamless integration of protein-specific knowledge with general language understanding, making it particularly suited for biological information retrieval tasks.
%

\subsection{Search Agent Architecture}

Our search agent architecture extends TourSynbio-7B's capabilities to address two critical limitations in protein engineering research: real-time access to protein information and automated tracking of scientific developments.
As depicted in Fig.~\ref{ex3}, the architecture comprises three main components: agent matching, parameter refinement, and execution, working in concert to deliver comprehensive search functionality.

Following the design of TourSynbio-Agent \cite{shen2024toursynbio}, the agent matching layer serves as the framework's initial interface, determining whether incoming queries require specialized agent intervention or can be handled directly by TourSynbio-7B's base capabilities. 
This layer employs designed classification prompts for each corresponding agent, such as ESMFold \cite{lin2023evolutionary} for protein structure prediction and Chroma for protein design \cite{ingraham2023illuminating}. 
Within the TourSynbio-Search, the agent referees the search agent, namely the PaperSearch agent or ProteinSearch agent.
The classification process utilizes carefully crafted prompts enhanced with chain-of-thought reasoning and few-shot learning prompting to accurately route queries to appropriate agents.

To ensure precise query routing, we implemented a hybrid classification approach combining predefined keywords with fuzzy matching capabilities. 
For instance, we recognize database-specific terms like ``UniProt'' and ``ArXiv'' while maintaining flexibility to understand variations in natural language queries. 
This dual approach improves classification accuracy while maintaining natural interaction patterns.
The agent matching process operates in parallel, evaluating multiple potential agent pathways simultaneously to determine the optimal response strategy. 
This parallel processing design enables rapid query assessment and ensures that complex queries requiring multiple agents' expertise are properly identified and routed. 
The TourSynbio-7B generates a prioritized list of relevant agents, as illustrated in Figure \ref{ex3}, enabling seamless coordination between different specialized components of the search framework.

\begin{figure*}[ht!]
\centering\centerline{\includegraphics[width=\linewidth]{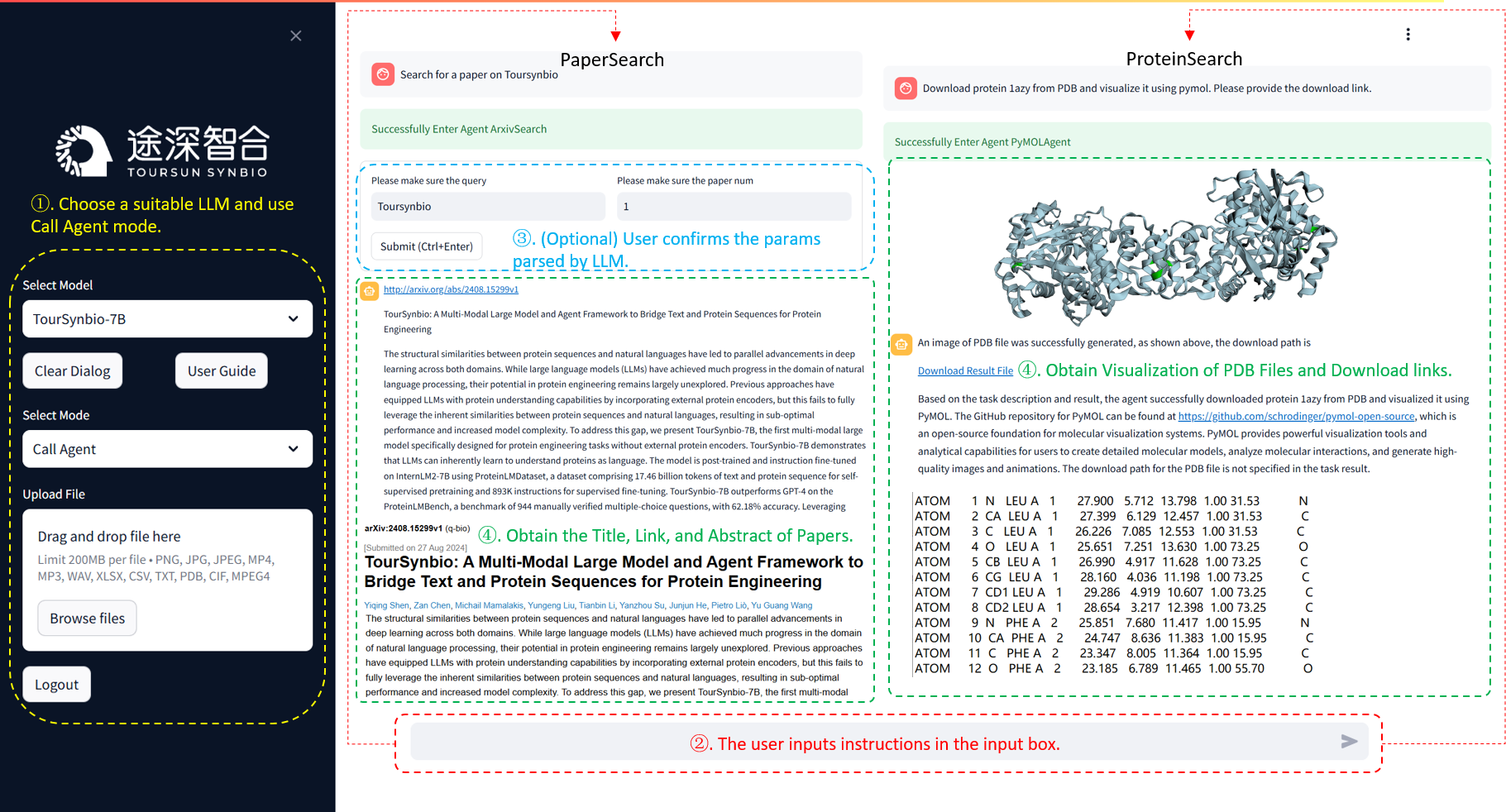}}
\caption{
Demonstration of the dual-module search interface in TourSynbio-Search. 
The interface is divided into three main components: (1) A model selection panel on the left that allows users to choose appropriate LLM configurations and agent modes, 
(2) PaperSearch agent in the center showing literature retrieval capabilities with parameter confirmation interface and search results for protein engineering papers, and (3) ProteinSearch agent on the right displaying integrated protein structure visualization and automated data retrieval features. 
The interface demonstrates the framework's unified approach to handling both literature and protein data queries while maintaining user control over search parameters and result presentation. 
The example shows a real-time search for ``TourSynbio''-related publications alongside protein structure analysis, illustrating the framework's capability to process multiple data types through its specialized agent architecture.
}
    \label{ex4}
\end{figure*}

\begin{tcolorbox}[
    colback=gray!10,  
    colframe=black,   
    boxrule=0.5pt,     
    label=prompt1,
    title=This is the Prompt used by the Toursynbio Search framework to determine whether to enter the Search Agent\, which utilizes techniques such as CoT and few shot to improve the accuracy of LLM parsing We have specified the analysis process of LLM and provided it with a specific example. Through this Prompt\, LLM will determine whether to call the Search Agent.
]
\label{prompt:classify}

You are a professional information extraction assistant capable of determining whether the user wishes to invoke a search function from the input. If so, return True; otherwise, return False.

If the user wishes to search, we should parse the following information from the user's input: PDB ID, Uniprot ID, Paper Keywords, ArXiv, BiorXiv.

During the analysis, please follow this thought process:

Identify the user's needs or tasks.

Based on the user's input, determine whether their needs are related to a search.

If the user's input indicates they do not want to use the search function or their intent cannot be analyzed, return False; otherwise, return True.

If the user's input does not provide the necessary information for performing a search, return False; otherwise, return True.

Here is an example for reference:

Q: Search for three papers related to CNN.

Thought process: The user wishes to search for papers related to 'Attention is all you need,' which is related to a search; we can clearly identify the user's intent and the specified number of papers to search for is three. Therefore, we determine that the user wishes to invoke the paper search function, returning True.

A: True

Now, please analyze the following inputs:

Q:
\end{tcolorbox}

If no specialized agents are identified as relevant (all `False' values in the classification output), the agent defaults to TourSynbio-7B's base dialogue capabilities.
Conversely, the detection of any `True' values triggers the specialized agent pathway, directing the query to the most appropriate expert agent based on the classification results.
Upon activation of the search agent (\textit{i}.\textit{e}., PaperSearch or ProteinSearch), TourSynbio-7B implements a parameter extraction protocol using carefully designed prompts. 
This extraction process identifies the necessary search parameters such as keywords, database preferences, and specific search constraints. 
The LLM automatically differentiates between paper-focused and protein-focused queries, routing them to either PaperSearch or ProteinSearch agents for specialized processing.

To optimize search efficiency and accuracy, we incorporated an interactive parameter refinement interface. 
This confirmation stage allows users to validate and modify the extracted search parameters before query execution. 
The interface presents the interpreted search parameters in a structured format, enabling users to fine-tune their queries and ensure alignment with their research objectives. 
This human-in-the-loop design reduces computational overhead while improving search precision by eliminating potential misinterpretations of user intent.

\begin{tcolorbox}[
    colback=gray!10,  
    colframe=black,   
    boxrule=0.5pt,     
    label=prompt2,
    title=This is the Prompt used by the Toursynbio Search framework to parse user input We have implemented different parsing strategy constraints for PaPerSearch and ProteinSearch Through this Prompt\, LLM will return the parsing results of the parameters in the user input.
]
\label{prompt:parse}

You are an information extraction assistant. You need to execute the following steps:

First, extract the search category the user wishes to use from their input, choosing between PaperSearch and ProteinSearch.

Next, if the user wants to use PaperSearch, you need to parse the search keywords (query) and the desired number of papers (paper\_num). Return the result in the format: "query: xxx, paper\_num: y, Mode: 'paper'". 

If the user wants to use ProteinSearch, extract the search keywords and return in the format: "query: xxx, Mode: 'protein'".

Example:
Q: Search for three papers related to CNN
A: query: 'CNN', paper\_num: 3, Mode: 'Paper'

Please extract information from the following sentence:
Q: {input\_text}

A: 

\end{tcolorbox}


\subsection{PaperSearch Agent}
The PaperSearch agent implements a specialized approach for extracting scientific literature from two primary preprint repositories, namely arXiv and bioRxiv. 
Due to differences in their API infrastructures, we developed distinct retrieval strategies for each database while maintaining a unified search interface for users.
For arXiv, we leverage their official API through the arXiv Python interface, enabling search functionality with multiple parameters. 
This implementation supports advanced query features through a flexible configuration. 
Users can perform keyword-based searching across multiple fields, customize result set sizes, configure search strategies, and specify various sorting options for result presentation, all through a streamlined interface that maintains consistency with the broader search framework.
The bioRxiv implementation required a more sophisticated approach due to platform constraints. 
Since bioRxiv's API only supports bulk paper downloads within specified timeframes and implements robust anti-scraping measures, we developed a custom database solution. 
This design maintains a repository of bioRxiv papers, updated through periodic synchronization with the platform's latest submissions. 
Our database architecture preserves essential metadata including titles, abstracts, keywords, and author information, enabling rapid and precise search capabilities.

\begin{figure}[htbp]
    \centering
    \includegraphics[width=1.0\linewidth]{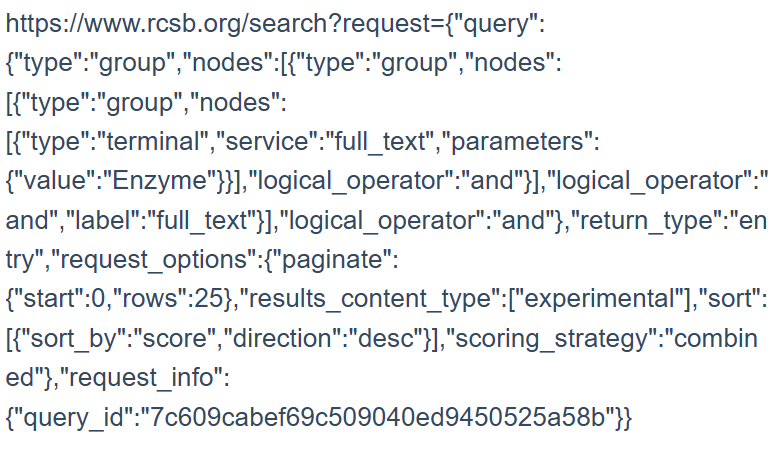}
    \caption{
    Example of a structured query URL for PDB database searches. This JSON-formatted request demonstrates the complex query structure required for protein searches, including search term specification (``Enzyme''), pagination parameters (25 rows per page), result type filtering (experimental data), and sorting criteria (score-based descending order)
    }
    \label{fig:enter-label}
\end{figure}

\subsection{ProteinSearch Agent}
The ProteinSearch agent implements two distinct retrieval pathways optimized for different search scenarios: direct identifier-based retrieval and keyword-based searching.
%
The identifier-based retrieval pathway processes explicit protein identifiers such as PDB IDs (\textit{e}.\textit{g}., ``\textit{1azy}'') or UniProt accession numbers (\textit{e}.\textit{g}., ``P0A2F2''). 
This agent leverages the standardized URL structures of major protein databases to generate direct download links. 
For PDB entries, the system constructs URLs following the pattern ``https://files.rcsb.org/download/{PDB\_ID}.pdb'', while UniProt entries are accessed via ``https://rest.uniprot.org/uniprotkb/{UniProt\_ID}.txt''. 
This standardized approach enables rapid, deterministic access to specific protein records.

For keyword-based searches, the agent employs a more sophisticated query construction. 
When searching the PDB database, our agent utilizes a comprehensive query template that incorporates up to 20 distinct search constraints, allowing for highly specific searches across multiple parameters. 
These constraints are exposed through an interactive confirmation dialog (shown in Fig.~\ref{ex4}), enabling users to refine their search criteria. 
The agent constructs specialized queries using the template ``https://www.rcsb.org/search?request=query\_list'', where query\_list encapsulates the user-specified constraints.

\section{Use Cases}
To validate the effectiveness and accessibility of our search framework, we present comprehensive demonstrations of both PaperSearch and ProteinSearch functionalities. These use cases illustrate how TourSynbio-Search simplifies complex database interactions for researchers across different expertise levels.

\subsection{Literature Retrieval Using PaperSearch}

\begin{figure}[!t]
\centerline{\includegraphics[width=1.0\linewidth]{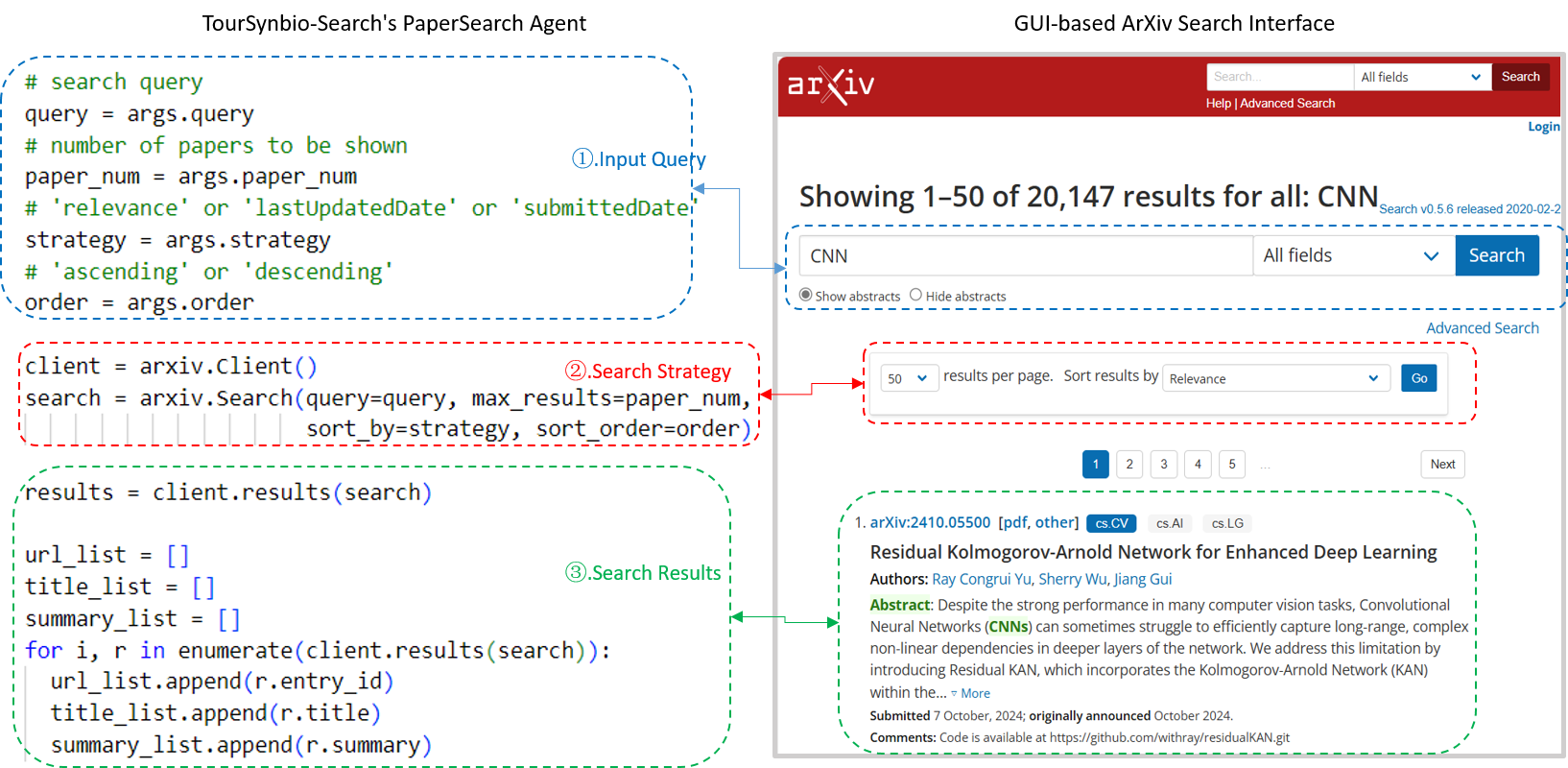}}
\caption{
Comparison between TourSynbio-Search's PaperSearch agent and traditional GUI-based arXiv search interface. 
The left panel illustrates the three-stage programmatic search implementation: (1) Query parameter handling with support for custom paper counts, relevance metrics, and sort orders, (2) Search execution using the arXiv API client with configurable result limits and sorting strategies, and (3) Structured result processing that extracts and organizes paper metadata including URLs, titles, and abstracts. 
The right panel shows the equivalent workflow in arXiv's web interface, demonstrating how TourSynbio-Search encapsulates complex API interactions behind a natural language interface while maintaining the full flexibility of programmatic search. 
This unified approach eliminates the need for users to toggle between different interfaces or understand API implementations while preserving advanced search capabilities.
}
\label{case1}
\end{figure}

We demonstrate the framework's literature retrieval capabilities through a comparative analysis between traditional GUI-based searches and our natural language interface. 
Using a representative scenario of searching for CNN-related publications, we showcase the framework's ability to transform simple queries into sophisticated database interactions.
When a user inputs a basic query such as ``CNN'' (illustrated in Fig.~\ref{case1}), our agent framework initiates a multi-stage processing pipeline. 
First, TourSynbio-7B analyzes the query to determine agent activation requirements. 
Upon confirming search agent engagement, TourSynbio-7B extracts key concepts and search parameters through natural language processing. 
These parameters undergo user verification before being transformed into structured database queries.

The framework's search execution layer implements distinct strategies for arXiv and bioRxiv databases. The core search functionality, shown in Fig.~\ref{case1}, supports a flexible parameter system with query content serving as the only required input. 
Additional parameters including result quantity, search strategy, and sort order are automatically configured by the LLM if not explicitly specified. 
By default, the agent returns three papers sorted in descending order of relevance, though these settings can be customized through natural language instructions.

The search results are presented in a streamlined interface, demonstrated in Fig. \ref{ex4}, that prioritizes essential information accessibility. Each result entry displays the paper's title, abstract, and direct access link, enabling researchers to efficiently evaluate content relevance without navigating away from the search interface. This integrated presentation eliminates the need for multiple browser windows or repeated searches across preprint repositories.
TourSynbio-Search's unified interface abstracts away the complexity of database-specific query syntax and API interactions, allowing researchers to focus on their scientific objectives rather than search mechanics. By consolidating all search operations within a single page and implementing natural language query processing, the framework reduces the cognitive overhead typically associated with literature searches.

\subsection{Protein Structure Retrieval and Visualization}

\begin{figure}[!t]
\centerline{\includegraphics[width=1.0\linewidth]{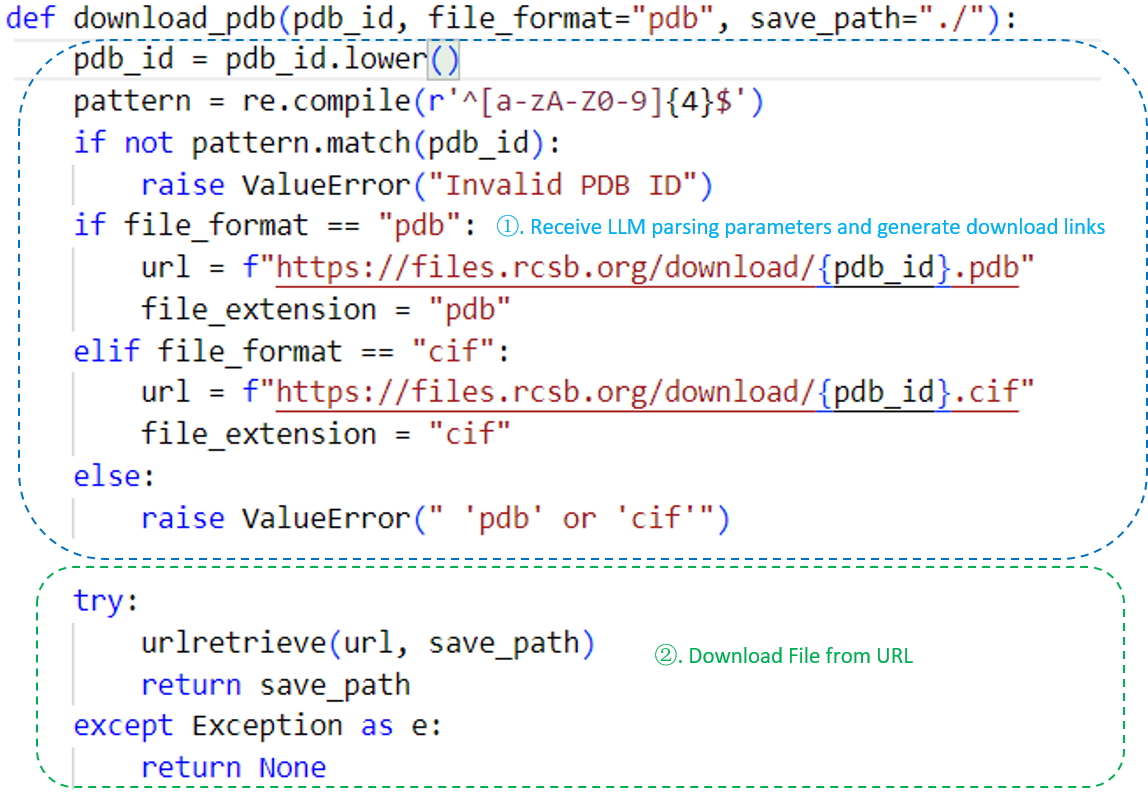}}
\caption{Implementation of the protein structure file retrieval module in TourSynbio-Search. The code demonstrates the two-stage download process: The first stage (blue section) handles input validation and URL generation, including PDB ID format verification using regular expressions and dynamic URL construction for both .pdb and .cif file formats. The second stage (green section) manages the actual file retrieval process through robust exception handling and standardized file saving procedures. This implementation ensures reliable access to protein structure data while maintaining compatibility with multiple structure file formats and providing graceful error handling for invalid requests.}
\label{case2}
\end{figure}

To demonstrate the framework's protein structure analysis capabilities, we present a case study of visualizing the three-dimensional structure of protein 1a2y from the PDB database (illustrated in Fig. \ref{ex4}). Users can initiate this complex workflow with a simple natural language request: ``\textit{Download protein 1a2y from PDB and visualize it using PyMOL. Also provide the download link for the PDB file}.''
The TourSynbio-Agent automatically identifies ``1a2y'' as a PDB identifier without requiring specific formatting or command syntax.

Upon user confirmation of the parsed parameters, TourSynbio-Search executes a multi-stage workflow that encompasses data retrieval, processing, and visualization. 
The agent first constructs the appropriate download URL for the PDB file, retrieves the structural data, and stores it in a designated directory. 
Following successful data acquisition, the framework activates its PyMOL visualization module, which renders the protein's three-dimensional structure according to preset visualization parameters. 
In addition to generating images, it also provides users with download links for PDB files, demonstrating that it can not only visualize but also facilitate data access and further analysis.

\section{Conclusion}

We present TourSynbio-Search, an innovative framework that bridges the gap between complex biological databases and user-friendly information retrieval. 
The framework demonstrates practical value in two aspects. 
For protein structure analysis, it enables direct retrieval and visualization of protein data through simple natural language queries. 
For literature research, it provides unified access to preprint servers and processes search results, streamlining the academic literature review process.
In brief, TourSynbio-Search offers researchers an efficient method to access and analyze protein-related data and literature. 

\bibliographystyle{IEEEtran}
\bibliography{main.bib}

\end{document}